\documentclass[twocolumn,showpacs,prb,floatfix,superscriptaddress]{revtex4}
\usepackage{graphicx}
\usepackage{bm}
\usepackage {amsmath}
\usepackage {amssymb}

\newcommand* {\vek}[1]{\bm{#1}}
\newcommand* {\abohr}{a_\mathrm{B}^\ast}
\newcommand* {\kk}{\bm{k}}
\newcommand* {\Ee}{\mathcal{E}}

\begin{document}

\title {Anomalous Spin Polarization of GaAs Two-Dimensional Hole
Systems}

\author{R. Winkler}
\affiliation {Institut f\"ur Festk\"orperphysik,
Universit\"{a}t Hannover, Appelstr.~2, D-30167 Hannover, Germany}
\affiliation{Department of Electrical Engineering,
Princeton University, Princeton, NJ 08544}
\author{E. Tutuc}
\author{S. J. Papadakis}
\author{S. Melinte}
\author{M. Shayegan}
\author{D. Wasserman}
\author{S. A. Lyon}
\affiliation{Department of Electrical Engineering,
Princeton University, Princeton, NJ 08544}

\date{\today}

\begin{abstract}
We report measurements and calculations of the spin-subband
depopulation, induced by a parallel magnetic field, of dilute GaAs
two-dimensional (2D) hole systems. The results reveal that the
shape of the confining potential dramatically affects the values
of in-plane magnetic field at which the upper spin subband is
depopulated. Most surprisingly, unlike 2D electron systems, the
carrier-carrier interaction in 2D hole systems does not
significantly enhance the spin susceptibility.
We interpret our findings using a multipole expansion of the spin
density matrix, and suggest that the suppression of the enhancement
is related to the holes' band structure and effective spin $j=3/2$.
\end{abstract}

\maketitle

\section{Introduction}

It was first pointed out by Janak that for a two-dimensional
electron system (2DES) in a static magnetic field the exchange
interaction acts like an effective magnetic field (in addition to
the applied field) so that the Zeeman energy splitting is enhanced.
\cite{jan69} Recently, the Zeeman splitting and spin susceptibility
of interacting 2D carrier systems have been a subject of renewed
interest, \cite{oka99, vit01, sha01, pud02, zhu03, tut03, shk04,
vak04, pru03, yoo99, pap00a, tut01, pro02, noh03, win04b, win04a,
zha05-cm} fueled by the promise of a paramagnetic to ferromagnetic
ground state transition at very low densities, \cite{blo29, sto38}
and the possibility that the spin polarization is related to the
apparent metal-insulator transition in dilute 2D systems.
\cite{abr01} Experiments have mostly focused on determining the spin
susceptibility from magneto-transport, \cite{oka99, vit01, sha01,
pud02, zhu03, tut03, shk04, vak04} and magnetization \cite{pru03}
measurements. The results generally show that the spin
susceptibility of 2DESs in different materials, e.g., Si,
\cite{oka99, vit01, sha01, pud02, pru03} GaAs, \cite{zhu03, tut03}
and AlAs \cite{shk04, vak04} increases as the density is reduced,
one report \cite{vit01} even suggesting a ferromagnetic instability
at the lowest densities.

Lately, the spin polarization of GaAs 2D \emph{hole} systems (2DHSs)
has become the subject of intensive research \cite{yoo99, pap00a,
tut01, pro02, noh03, win04b, win04a} because the holes have a larger
effective mass (than electrons) so that they can be made effectively
more dilute while maintaining high quality. Furthermore, the spin
polarization of holes is important in the context of ferromagnetic
semiconductors such as GaMnAs where it is known that the
ferromagnetism is mediated by the itinerant valence band holes.
\cite{die01, abo01} We show here that the spin susceptibility of
2DHSs depends dramatically on the shape of the confining potential.
Moreover, we find that, in contrast to their 2D electron
counterparts, dilute 2DHSs exhibit no significant enhancement of the
spin susceptibility as compared with calculations which neglect
exchange-correlation. \cite{mass} We will argue that this surprising
behavior is related to the holes' band structure and the fact they
have effective spin $j=3/2$ rather than $j=1/2$ which is the case
for electrons.

\section{Sample parameters and experimental details}

Four samples from different wafers, including two GaAs/AlGaAs
heterojunctions and two GaAs quantum wells (QWs) flanked by AlGaAs
barriers, were investigated in this study (Table~\ref{tab:samples}).
Depending on their substrate orientation and carrier type, our
samples were either Be-doped (samples H, Q1) or Si-doped (Q2, A).
All samples were fitted with metal front and back gates to control
their density as well as the electric field perpendicular to the 2D
systems. We made measurements in $^{3}$He or dilution refrigerators
down to a temperature $T = 0.03$~K and in magnetic fields up to
25~T.

\begin{table}[tbp]
  \caption{\label{tab:samples}Typical
  densities  $n$ (in $10^{10}$~cm$^{-2}$) and mobilities $\mu$
  (in m$^2$/Vs) of the samples used in this study.}
  \tabcolsep 0.5em
  \begin{tabular}{*{6}{c}} \hline\hline
    sample & carriers & structure & substrate & $n$ & $\mu$ \\ \hline
    H  & holes & heterojunction    & (001) & $5.3$ & $30$ \\
    Q1 & holes & 150~{\AA} wide QW & (001) & $4.8$ & $11$ \\
    Q2 & holes & 200~{\AA} wide QW & (113)A& $6.8$ & $55^\dagger$ \\
    A  & electrons & heterojunction & (001) & $3.0$ & $48$
    \\ \hline \hline
  \end{tabular}
  {$^\dagger$}$\mu$ for $\vek{I} \parallel [33\overline{2}]$.
  For $\vek{I} \parallel [\overline{1}10]$ we have $\mu =
  35$~m$^2$/Vs.
\end{table}

\section{Experimental results for (001) 2D holes}

\begin{figure}[tbp]
  \includegraphics[width=1.0\columnwidth]{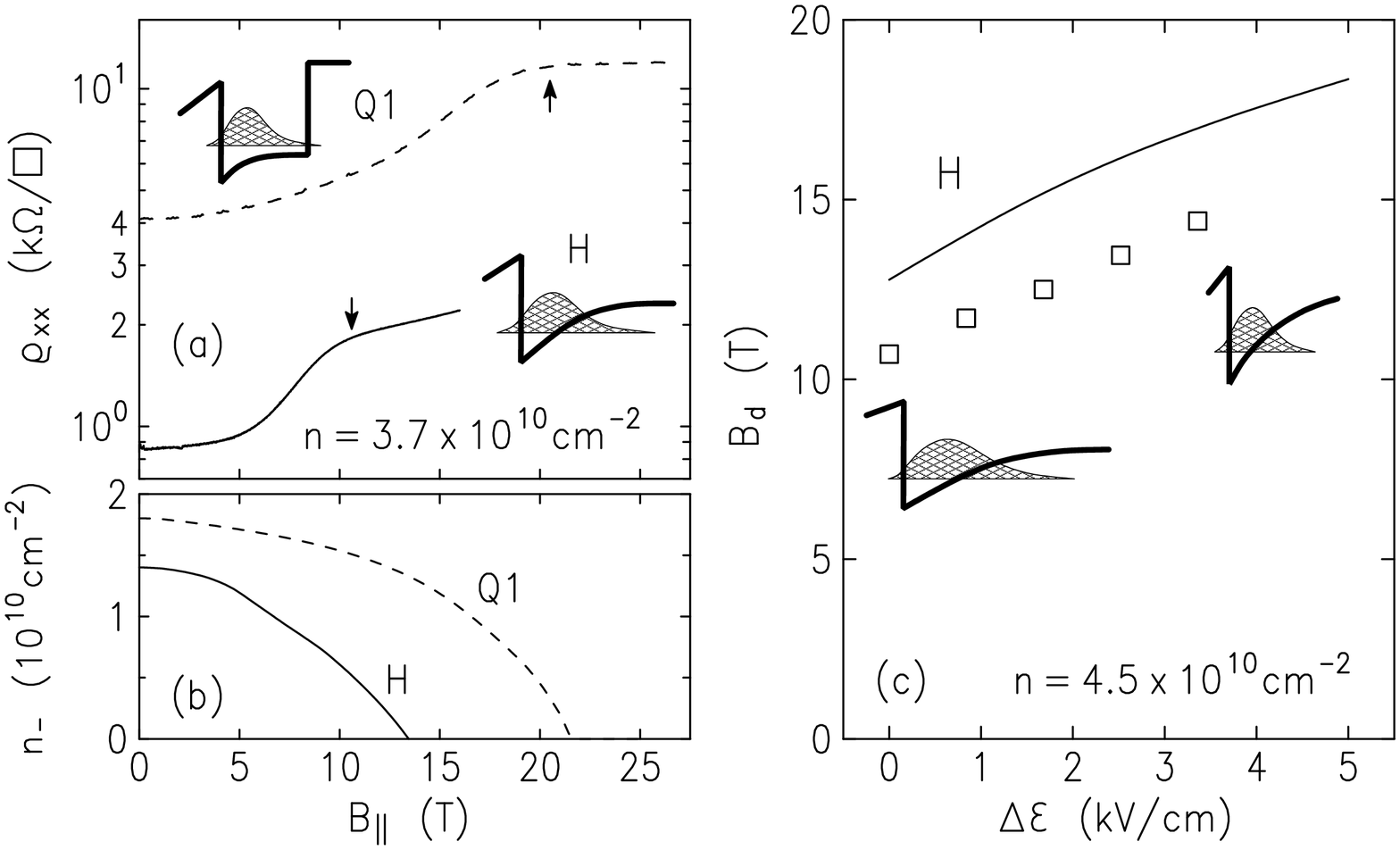}
  \caption{\label{fig:het-qw}(a) Longitudinal resistivity
  $\rho_{xx}$ versus in-plane magnetic field $B_\|$ measured at $T =
  0.3$~K for 2D hole samples, H and Q1, at the same density $n =
  3.7 \times 10^{10}$~cm$^{-2}$. The depopulation fields $B_d$ are
  marked by arrows.  (b) Calculated density $n_-$ in the minority
  spin subband of samples H and Q1 as a function of $B_\|$.  (c)
  Measured (squares) and calculated (solid line) depopulation field
  $B_d$ versus change $\Delta \Ee$ of the electric field in sample H
  for constant density $n = 4.5 \times 10^{10}$~cm$^{-2}$.}
\end{figure}

\begin{figure}[tbp]
  \includegraphics[width=1.0\columnwidth]{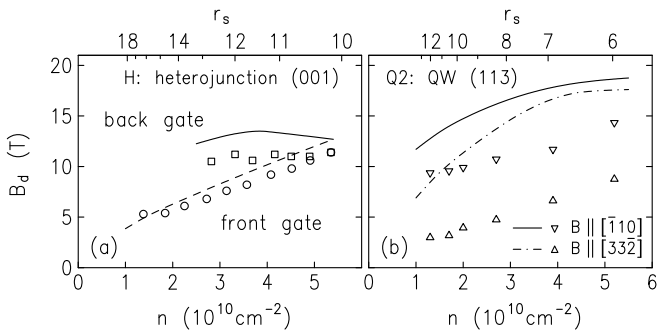}
  \caption{\label{fig:bd}Measured (symbols) and calculated (lines)
  depopulation field $B_d$ as a function of $n$ for samples (a) H
  and (b) Q2. In (a), for the solid line and squares (dashed line
  and circles) $n$ was varied via a back (front) gate.  In (b) the
  different symbols refer to $\vek{B} \parallel [\overline{1}10]$
  and $\vek{B} \parallel [33\overline{2}]$ as indicated.  The upper
  horizontal axes show the calculated density paramater $r_s$ for
  the corresponding $n$. In (a) we used $r_s (n)$ for the front
  gate.}
\end{figure}

In Fig.~\ref{fig:het-qw}(a) we show the longitudinal resistivity
$\rho_{xx}$ versus in-plane magnetic field $B_\|$ for samples H and
Q1 both measured at a density of $n = 3.7 \times 10^{10}$~cm$^{-2}$.
The data shows a positive magnetoresistance with a marked change in
functional form above the magnetic field $B_{d}$ that reflects the
complete depopulation of the minority spin subband. \cite{oka99,
tut01} In Fig.~\ref{fig:het-qw}(a) $B_d$ is marked by arrows.
Remarkably, the field $B_{d}$ depends greatly on the shape of the
confining potential. Indeed, we have $B_{d} \simeq 10.6$~T for
sample H and $B_{d} \simeq 20.5$~T for sample Q1, even though the
data were taken at the {\it same} density.
In Fig.~\ref{fig:het-qw}(c) we show $B_d$ in sample H when the
electric field $\Ee$ across the junction is varied by means of front
and back gates such that $n$ is kept constant at $4.5 \times
10^{10}$~cm$^{-2}$. The field $B_d$ increases significantly with
increasing $\Ee$.
In Fig.~\ref{fig:bd}(a) we show the measured $B_d$ versus $n$ for
sample H. The values of $B_d$ depend rather sensitively on whether
$n$ is changed by means of a front or back gate.

\section{Calculations and discussions}

In order to explain the experimental results of
Figs.~\ref{fig:het-qw} and~\ref{fig:bd} we have performed
parameter-free calculations in the multiband envelope-function and
self-consistent Hartree approximations for the quasi-2D system.
\cite{win03b, tut03} Figure \ref{fig:het-qw}(b) shows the calculated
density $n_-$ in the minority spin subband as a function of $B_\|$.
The lines in Figs.~\ref{fig:het-qw}(c) and~\ref{fig:bd} show the
calculated $B_d$ for the corresponding experiments. \cite{gate} The
calculations reproduce the different behavior of samples H and Q1 in
satisfactory agreement with experiment.

\subsection{Confinement potential dependence of $B_d$}

As we discuss in the next section, the close agreement between the
experimental and calculated $B_d$ in Figs.~\ref{fig:het-qw}
and~\ref{fig:bd}(a) is very surprising because the calculations do
not take exchange-correlation effects into account.  Such effects
are indeed dominant for 2D \emph{electron} systems \cite{jan69,
oka99, vit01, sha01, pud02, zhu03, tut03, shk04, vak04, pru03,
raj77, tan89, att02, zha05-cm} that are as dilute as the 2DHSs of
Figs.~\ref{fig:het-qw} and~\ref{fig:bd}(a). Before elaborating on
this aspect of our results, however, we first discuss the remarkably
strong dependence of $B_d$ on the confining potential. For an ideal,
strictly 2D system with effective mass $m^\ast$ and effective $g$
factor $g^\ast$ we have $B_d \propto 1 / (m^\ast g^\ast)$
independent of the shape of the confining potential. To understand
the surprising results in Figs.~\ref{fig:het-qw} and~\ref{fig:bd} we
will first concentrate on the Zeeman splitting which gives rise to
the dominant contribution of the confinement dependence of $B_d$ in
2DHSs. Then we discuss the effect of $B_\|$ on the orbital motion.

Unlike electrons in the conduction band that have spin $1/2$, holes
in the uppermost valence band are characterized by an effective spin
$3/2$ (Ref.~\onlinecite{win03b}). Subband quantization in 2DHSs
yields a quantization of angular momentum with $z$ component $m =
\pm 3/2$ for the heavy holes (HHs) and $m = \pm 1/2$ for the light
holes (LHs). In our samples, only the lowest HH subband is occupied.
The quantization axis of angular momentum that is enforced by HH-LH
splitting points perpendicular to the 2D plane. The Zeeman energy
splitting due to $B_\|$ thus competes with the HH-LH splitting and
it is well-known that the $B_\|$-linear Zeeman splitting of HH
states is suppressed. \cite{kes90, lin91} [The simple model of
Ref.~\onlinecite{kes90} yields $B_d \simeq 250$~T for the systems in
Fig.~\ref{fig:het-qw}(a).] In the following we will discuss why the
depopulation fields $B_d$ observed in real 2DHSs are much smaller
than what these arguments suggest.

The dispersion of HH states is known to be highly nonparabolic as a
consequence of HH-LH coupling. \cite{win03b} Therefore, the
suppression of Zeeman splitting linear in $B_\|$ is merely the
lowest-order effect in a Taylor expansion of the spin-split
dispersion $E_\sigma (\kk_\|, B_\|)$ of HH states as a function of
the (canonical) wave vector $\kk_\|$, $B_\|$, and spin index
$\sigma$. Mixed higher-order terms proportional to $B_\|$ and
$\kk_\|$ give rise to an average Zeeman splitting of the occupied
hole states which is approximately linear in $B_\|$. Thus we find
that $B_d$ is generally much smaller than the value one would expect
if the $B_\|$-linear Zeeman splitting were suppressed. This is also
consistent with previous experimental data for 2DHSs that were
interpreted ignoring completely the suppression of $B_\|$-linear
Zeeman splitting in HH systems. \cite{yoo99, pap00a, tut01, pro02,
noh03}

Now we can understand why the Zeeman energy splitting in 2DHSs
depends sensitively on the shape of the confining potential. The
mixed higher-order terms that are responsible for the Zeeman energy
splitting $E_Z (B_\|)$ of HH systems compete with the HH-LH
splitting. The latter depends sensitively on the shape of the
confining potential so that we have here a tool to tune $E_Z (B_\|)$
of 2DHSs. In narrow quasi-2D HH systems we have a large HH-LH
splitting so that the Zeeman energy splitting is reduced, giving
rise to a large $B_d$. We get a large $E_Z (B_\|)$ (a small $B_d$)
in wide systems. We can define $E_Z (B_d)$ as the energy difference
between the Fermi energy and the subband edge at $B_d$. In the wide
heterojunction of Fig.~\ref{fig:het-qw}(b), the calculated $E_Z
(B_d)$ is $0.44$~meV, significantly larger than $E_Z (B_d) =
0.26$~meV in the narrower QW, despite the smaller value of $B_d$ in
the heterojunction. Similarly, the increase of $B_d$ with increasing
$\Delta \Ee$ in Fig.~\ref{fig:het-qw}(c) reflects the change of the
HH-LH splitting in the system.

Next we discuss the effect of $B_\|$ on the orbital motion. In
general, \cite{ste68} the mass of the particles in quasi-2D systems
increases as a function of $B_\|$ which reflects the fact that,
ultimately, for large $B_\|$ resulting in a magnetic length
comparable to the width of the quasi-2D system the particle states
become dispersionless Landau levels. Obviously, this effect depends
on the thickness of the quasi-2D system and it has been shown that
$B_d$ in wide quasi-2D \emph{electron} systems is much smaller than
$B_d$ in narrow 2DESs. \cite{tut03} We will argue next that the mass
enhancement does not explain, however, the results in
Figs.~\ref{fig:het-qw} and~\ref{fig:bd}(a).

Our numerical calculations show, in agreement with the 2DESs'
results, \cite{tut03} that the mass enhancement at small $B_\|$ is
smaller in the QW than in the heterojunction. However, $m^\ast$ in
2DHSs increases highly nonlinearly as a function of $B_\|$ which is
particularly important for the QW with the larger $B_d$. Thus we
find that at $B_d$ the mass enhancement in the narrower 2DHS of the
QW is larger than in the wide 2DHS of the heterojunction. We note
that at $B_d$ the mean kinetic energy equals approximately half the
Zeeman energy splitting $E_Z (B_d)$ so that for
Fig.~\ref{fig:het-qw}(b) the mass enhancement can be inferred from
the $E_Z$ values quoted above [see also Eq.\ (\ref{eq:avmass})
below].

The anomalous enhancement of $m^\ast$ at $B_d$ with decreasing width
of the quasi-2D HH system depends sensitively on the system
parameters such as the density and the shape of the confining
potential. For the paramteres in Fig.~\ref{fig:het-qw}(c), $m^\ast$
at $B_d$ is approximately independent of $\Delta \Ee$ (despite the
significant change of $B_d$), i.e., the increase of $B_d$ with
$\Delta \Ee$ is essentially only due to the decrease of the Zeeman
splitting discussed above. For about twice the largest field $\Delta
\Ee$ we could reach experimentally one enters the regime when
$m^\ast$ at $B_d$ starts to increase with $\Delta \Ee$.

\subsection{Lack of spin susceptibility enhancement for (001) 2D holes}

\begin{figure}[tbp]
  \includegraphics[width=0.6\columnwidth]{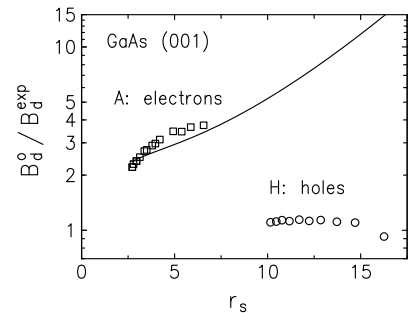}
  \caption{\label{fig:elrat}Ratio $B_d^0 / B_d^\mathrm{exp}$ of the
  depopulation field $B_d^0$ calculated neglecting
  exchange-correlation to the measured field $B_d^\mathrm{exp}$ for
  a 2DES (squares) and a 2DHS (circles) in GaAs (001)
  heterojunctions plotted versus the density parameter $r_s$. In
  both cases, $n$ was varied via a front gate.
  For electrons, the solid line shows the ratio $B_d^0 /
  B_d^\mathrm{xc}$, where $B_d^\mathrm{xc}$ was calculated taking
  into account exchange-correlation. \cite{tut03, att02}}
\end{figure}

A most remarkable aspect of the results in Figs.~\ref{fig:het-qw}
and~\ref{fig:bd}(a) is the reasonable \emph{quantitative} agreement
between the experimental data and the calculations. This is
particularly puzzling because many-particle effects beyond the
Hartree approximation (i.e., exchange-correlation effects) were not
taken into account. This is in sharp contrast to the case of dilute
2DESs for which it is known that exchange-correlation significantly
increases the spin susceptibility when $n$ is reduced. \cite{jan69,
oka99, vit01, sha01, pud02, zhu03, tut03, shk04, vak04, pru03,
raj77, tan89, att02, zha05-cm} To quantify this point we show in
Fig.~\ref{fig:elrat} the ratio $B_d^0 / B_d^\mathrm{exp}$ of the
depopulation field $B_d^0$, calculated neglecting
exchange-correlation, to the experimentally measured field
$B_d^\mathrm{exp}$ for a 2DES (squares) \cite{sampleA} and the 2DHS
(circles) in sample H. The ratio thus reflects the enhancement of
the spin susceptibility at $B_d$ due to exchange-correlation. Our
results are plotted as a function of the dimensionless density
parameter $r_s$ defined as the average interparticle spacing
measured in units of the effective Bohr radius $\abohr$, $r_s \equiv
1 / (\abohr \sqrt{\pi n})$.

For the 2DES in Fig.~\ref{fig:elrat} the ratio $B_d^0 /
B_d^\mathrm{exp}$ is between 2 and 4 and, as expected, it increases
with $r_s$. We also remark that for electrons the experimentally
observed reduction of $B_d$ is in reasonable quantitative agreement
with numerical calculations that take exchange-correlation into
account. To illustrate this point, the solid line in
Fig.~\ref{fig:elrat} shows the ratio $B_d^0 / B_d^\mathrm{xc}$,
where $B_d^\mathrm{xc}$ was calculated in the framework of
spin-density-functional theory using a parameterization of the
polarization-dependent exchange-correlation potential that was
recently obtained by means of quantum Monte Carlo calculations.
\cite{tut03, att02} For the 2DHS, on the other hand, the expected
enhancement of the spin susceptibility and $B_d^0 /
B_d^\mathrm{exp}$ ratio is conspicuously absent in
Fig.~\ref{fig:elrat}.  Note that, because of their larger effective
mass compared to GaAs electrons ($m^\ast \simeq 0.25$ compared to
$m^\ast = 0.067$; here $m^\ast$ is given in units of the
free-electron mass), 2DHSs have significantly larger $r_s$ and are
thus effectively much more dilute. Nonetheless, the ratio $B_d^0 /
B_d^\mathrm{exp}$ remains close to unity up to the largest values of
$r_s$ where a greater than ten-fold enhancement is expected.

\begin{figure}[tbp]
  \includegraphics[width=1.0\columnwidth]{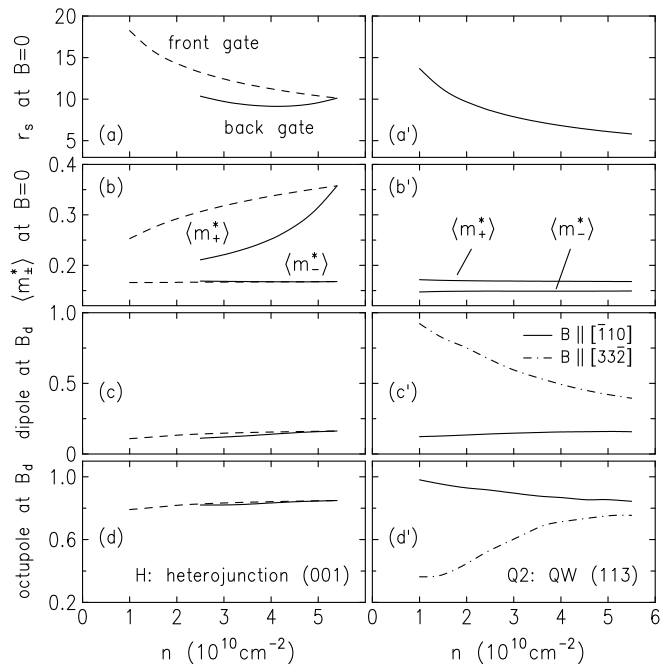}
  \caption{\label{fig:bdata}Calculated density parameter $r_s$,
  average effective mass $\langle m^\ast_\pm \rangle$, normalized
  dipole and octupole moments \cite{multipol} versus $n$ for samples
  H and Q2.  Line styles have the same meaning as in
  Fig.~\ref{fig:bd}.}
\end{figure}

Before discussing possible reasons for this anomalous behavior of
2DHSs, we make remarks regarding the effective mass $m^\ast$ which
enters $\abohr$ and thus determines $r_s$. For holes, $m^\ast$ is
not uniquely defined. As discussed above, the HH dispersions are
typically nonparabolic, meaning that $m^\ast$ depends on energy and
therefore on $n$ and the confinement potential. Moreover, the HH
systems have a large Rashba and Dresselhaus spin splitting at $B =
0$ (Ref.~\onlinecite{win03b}), leading to two energy versus wavevector
($\kk_\|$) dispersions with different curvatures and effective
masses $m^\ast_+$ and $m^\ast_-$. Commonly, values of $m^\ast$
between about 0.2 and 0.4 are used for holes in GaAs. \cite{yoo99,
pro02, noh03, pan03, win03b} Here we adopt a simple definition for
an average effective mass $\langle m^\ast \rangle$:
\begin{equation}
  \label{eq:avmass}
  \langle m^\ast \rangle = \hbar^2\pi n / (2 \langle E_k \rangle) \; ,
\end{equation}
where $\langle E_k \rangle$ is the mean kinetic energy per particle.
Figure~\ref{fig:bdata}(a) shows the calculated density parameter
$r_s$ in sample H, when $n$ is changed by means of a front or back
gate. Note that for a single, parabolic dispersion with an effective
mass $m^\ast$, the mass $\langle m^\ast \rangle$ as defined in Eq.\ 
(\ref{eq:avmass}) properly reduces to $m^\ast$ and is independent of
$n$. For the 2DHS, on the other hand, $\langle m^\ast \rangle$ in
general depends sensitively not only on $n$ but also on the system's
parameters such as the thickness of the 2DHS and on the applied
electric and magnetic fields, as discussed above.
If we take into account Rashba and Dresselhaus spin splitting,
\cite{win03b} then we get, similar to Eq.\ (\ref{eq:avmass}),
effective masses $\langle m^\ast_\pm \rangle$ for each spin subband.
To illustrate this effect, we show $\langle m^\ast_\pm \rangle$ for
sample H in Fig.~\ref{fig:bdata}(b).
We emphasize that the main conclusion of our work, namely the lack
of enhancement of the spin susceptibility with increasing
diluteness, is not affected by the specific values of $m^\ast$ used
to define $r_s$: it is clear in Fig.~\ref{fig:elrat} that if $r_s$
were changed by a factor of 2 or 3, there would still exist a large
discrepancy between the experimental hole data and the expected
enhancement.

Why do dilute 2DHSs \emph{not} show a significant enhancement of the
spin susceptibility? Using a recently developed multipole expansion
of the spin density matrix \cite{win04b} we argue in the remainder
of this paper that the $j=3/2$ hole spin is the likely culprit.
For 2DESs with spin $1/2$ it is well-known that the mean Coulomb
energy $\langle E_c \rangle$ per particle can be completely
characterized using $n$ and (the magnitude of) the spin polarization
$\vek{\zeta}$ as independent parameters. \cite{raj77} This is
because the $2 \times 2$ spin density matrix of spin $1/2$ systems
can be decomposed into four independent terms: $n$ (a monopole) and
the three components of the spin polarization vector~$\vek{\zeta}$
(a dipole). \cite{win04b} In the Hartree-Fock (HF) approximation,
the direct part of the Coulomb energy $\langle E_c^\mathrm{HF}
\rangle$ cancels the potential of the positive background so that
only the exchange term $\langle E_x \rangle$ remains, \cite{kit63,
ste73}
\begin{equation}
  \label{eq:ex}
  \langle E_c^\mathrm{HF} \rangle =
  \langle E_x \rangle
  = - \frac{2 e^2}{3\pi}
    \sqrt{2\pi n}
    \left[(1+\zeta)^{3/2} + (1-\zeta)^{3/2}\right] .
\end{equation}
The Coulomb energy $\langle E_c^\mathrm{HF} \rangle$ is thus
proportional to $\sqrt{n}$.  Higher order terms in a series
expansion for $\langle E_c \rangle (n, \zeta)$ of 2DESs were
calculated in Ref.~\onlinecite{raj77}. $\langle E_c \rangle$ was
calculated numerically in, e.g., Refs.\ \onlinecite{tan89}
and~\onlinecite{att02}.

The $4\times 4$ spin density matrix of $j=3/2$ 2DHSs, on the other
hand, can be decomposed into four multipoles, where the monopole is
the density $n$, the dipole corresponds to the spin polarization at
$B>0$, the quadrupole reflects the HH-LH splitting, and the octupole
is a unique feature of $j=3/2$ hole systems at $B>0$
(Ref.~\onlinecite{win04b}). For sample H, the normalized dipole and
octupole at $B_d$ are shown in Figs.~\ref{fig:bdata}(c) and (d)
(Ref.~\onlinecite{multipol}). Unlike 2DESs, the dipole at $B_d$ is
much smaller than unity, i.e., despite the fact that only one spin
subband is occupied at $B_d$, the system is only weakly spin
polarized. This result \cite{win04a} is an immediate consequence of
the suppression of $B_\|$-linear Zeeman splitting \cite{kes90,
lin91} discussed above. The octupole can be interpreted as a new
``spin degree of freedom'' of spin $3/2$ hole systems at $B>0$ which
does not exist for the more familiar case of spin $1/2$ electron
systems. When the spin polarization is suppressed for an in-plane
magnetic field, the 2D HH systems aquire instead a large octupole
moment, \cite{either} as visible in Fig.~\ref{fig:bdata}(d). The
quadrupole is always close to unity because the HH-LH mixing is
small in the systems considered here. Therefore, the quadrupole is
not shown in Fig.~\ref{fig:bdata}. By definition, these four
multipoles provide a set of \emph{independent} parameters that can
be used to parameterize the Coulomb energy $\langle E_c \rangle$ of
spin $3/2$ systems, similar to $\langle E_c \rangle (n, \zeta)$ in
spin $1/2$ systems. \cite{fermi} However, the series expansion is
presently not known and its calculation represents a formidable
task. Our study indicates that the series expansion of $\langle E_c
\rangle$ of spin $3/2$ 2DHSs is qualitatively different from
$\langle E_c \rangle (n, \zeta)$ of spin $1/2$ 2DESs.

The HF exchange energy $\langle E_x \rangle$ of 2D HH systems at
$B=0$ is the same as $\langle E_x \rangle$ of spin $1/2$ 2DESs
because Eq.\ (\ref{eq:ex}) requires only that the eigenstates of the
two spin subbands for the same $\kk_\|$ are orthogonal. \cite{kit63}
For a HH system, the main effect of a perpendicular magnetic field
$B_\perp$ is a spin polarization (a dipole), whereas an in-plane
field $B_\|$ usually gives rise to an octupole moment. \cite{win04b,
either} The spin density matrices of 2D HH systems at $B_\perp > 0$
and $B_\| >0$ are thus qualitatively different. However, the HF
exchange energy does not distinguish between these cases and always
leads to the same enhancement of the exchange energy as in 2DESs. We
note that different results can be obtained for $\langle E_x
\rangle$ when HH-LH mixing is significant. \cite{macd-priv} Also,
different results are obtained for $\langle E_c \rangle$ in
higher-order perturbation theory when the more complicated energy
dispersion must be taken into account. These are the reasons why the
well-established results for exchange-correlation in dilute spin
$1/2$ 2DESs cannot easily be transferred to spin $3/2$ 2DHSs.

\section{Results for (113) 2D holes}

We extend our investigation by comparing the results for sample H
with the data for Q2, a QW grown on a (113)A GaAs substrate.
Figure~\ref{fig:bd}(b) shows $B_d$ versus $n$ for Q2. The field
$B_d$ strongly depends on whether $\vek{B}_\|$ is applied in the
in-plane crystallographic directions $[\overline{1}10]$ or
$[33\overline{2}]$ (Ref.~\onlinecite{sampleQ2}). The right column of
Fig.~\ref{fig:bdata} shows $r_s$, $\langle m^\ast_\pm \rangle$, the
dipole and the octupole moments calculated for Q2.
For this sample, when $\vek{B}_\|$ is applied parallel to
$[33\overline{2}]$, the measured $B_d$ is well below the calculated
value. It is remarkable that for this particular geometry, the
octupole remains small, but the 2DHS develops a large dipole moment
[Figs.~\ref{fig:bdata}(c') and (d')], similar to 2DESs in a
$\vek{B}_\|$ (Ref.~\onlinecite{b332}). This observation suggests
that the spin susceptibility is enhanced by many-particle effects
only when the magnetic field gives rise to a spin polarization. On
the other hand, Figs.~\ref{fig:bd} and~\ref{fig:bdata} suggest that
there is no significant enhancement in $j=3/2$ 2DHSs with a large
octupole but a small dipole moment.

\section{Summary and conclusions}

We have shown that the spin susceptebility of dilute GaAs 2DHSs in
an in-plane magnetic field $B_\|$ depends sensitively on the shape
of the confining potential. Most remarkably, the spin susceptibility
is not significantly enhanced as compared with calculations which
neglect the carrier-carrier interaction. This is in sharp contrast
to dilute electron systems for which it is known that many-body
effects greatly enhance the spin susceptibilty. Using a multipole
expansion of the spin density matrix we have argued that the
suppression of the enhancement is related to the holes' band
structure and effective spin $j=3/2$.

Our findings have important implications for the quantum phase
diagram of dilute 2DHSs. In dilute \emph{electron} systems, the
exchange-correlation enhancement of the spin susceptibility can be
considered a precursor for the ferromagnetic liquid which is
expected to be the ground state of ultra-low density 2DESs with $r_s
\gtrsim 26$ (Ref.~\onlinecite{att02}). The extra multipoles of 2DHSs
provide new possibilities for the ground state of hole systems to
respond to external perturbations such as a magnetic field thus
leading to a richer phase diagram than in spin $1/2$ electron
systems. \cite{fermi} However, our results suggest that a
ferromagnetic phase (i.e., a fully spin-polarized phase with a
maximum dipole moment) is often not favored in dilute 2DHSs. This
could also have important implications for ferromagnetic
semiconductors such as GaMnAs where it is known that the
ferromagnetism is mediated by the itinerant spin $3/2$ holes in the
valence band. \cite{die01, abo01} In itinerant ferromagnets it is
the polarization-dependent competition between the Coulomb energy
and the kinetic energy of the interacting carriers which controls
the ferromagnetic transition.


\begin{acknowledgments}
  We thank BMBF, DOE, NSF, and ARO for support, and D.~C.\ Tsui and
  A.~H.\ MacDonald for illuminating discussions. Part of this work
  was done at NHMFL; we thank G.\ Armstrong, T.\ Murphy, and E.\ 
  Palm for support.
\end{acknowledgments}



\begin{thebibliography}{42}
\bibitem[{gat()}]{gate}
\bibinfo{note}{In the calculations, front and back gate are modeled by
  different boundary conditions for the Hartree potential.}
\bibitem[{sam({\natexlab{a}})}]{sampleA}
\bibinfo{note}{Electron data in Fig.~\ref{fig:elrat} are from sample A of
  Ref.~\onlinecite{tut03}.}
\bibitem[{mul()}]{multipol}
\bibinfo{note}{The multipoles are normalized such that the largest possible
  values are unity.}
\bibitem[{sam({\natexlab{b}})}]{sampleQ2}
\bibinfo{note}{Experimental data for sample Q2 are from
  Ref.~\onlinecite{pap00a}. Within experimental accuracy, they are
  independendent of the direction of the current.}
\bibitem[{b33()}]{b332}
\bibinfo{note}{The large dipole of 2DHSs in an in-plane magnetic field
  $B_{[33\overline{2}]}$ is a consequence of the low symmetry of the (113)
  surface which results in a highly anisotropic Zeeman splitting of the 2DHS.
  Within the $2\times 2$ subspace of HH states, the in-plane field
  $B_{[33\overline{2}]}$ gives rise to an off-diagonal Zeeman term $\propto
  B_{[33\overline{2}]} \,\sigma_z$, where $\sigma_z$ is the Pauli spin matrix
  for the $z$ direction perpendicular to the 2D plane. \cite{win03b} So in this
  particular case, the in-plane field acts like a perpendicular field.}
\end{thebibliography}
\end{document}